\documentclass[psamsfonts]{amsart}

\usepackage{natbib}
\numberwithin{equation}{section}

\pdfoutput=1
\usepackage{graphics}
\usepackage[dvipdf]{graphicx}
\usepackage{epstopdf}
\usepackage{epsfig}
\usepackage{url}

\newcommand \be{\begin{equation}}
\newcommand \bea{\begin{eqnarray}}
\newcommand \ee{\end{equation}}
\newcommand \eea{\end{eqnarray}}

\usepackage{epsf}

\usepackage{subfigure}
\usepackage{color}

\begin{document}
\title[Icequakes and precursors]{Icequakes as precursors of ice avalanches}

\author{J. Faillettaz}
\address{VAW, ETH Zurich, Laboratory of Hydraulics, Hydrology and Glaciology\\Switzerland} 
\email{faillettaz@vaw.baug.ethz.ch}
\url{http://www.glaciology.ethz.ch}
\author{M. Funk}
\address{VAW, ETH Zurich, Laboratory of Hydraulics, Hydrology and Glaciology\\Switzerland}
\email{funk@vaw.baug.ethz.ch}
\author{ D. Sornette}
\address{Department of Management, Technology and Economics, ETH Z\"urich}
\address{Department of Earth Sciences, ETH Z\"urich}
\address{Institute of Geophysics and Planetary Physics, UCLA}
\curraddr{Department of Management, Technology and Economics,\\ ETH Z\"urich\\Switzerland}
\email{ dsornette@ethz.ch}



\date{}



\begin{abstract}
A hanging glacier at the east face of Weisshorn broke off in 2005. We were
able to monitor and measure surface motion and icequake activity for 21 days up to
three days prior to the break-off. Results are presented from the analysis of
seismic waves generated by the glacier during the rupture maturation process. 
Three types of precursory signals of the imminent catastrophic rupture were identified: 
(i) an increasing seismic activity within the glacier, (ii)
a change in the size-frequency distribution of icequake energy, and
(iii) a log-periodic oscillating behavior superimposed on power law
acceleration of the inverse of waiting time between two icequakes. 
The analysis of the seismic activity gave indications of the rupture
process and led to the identification of two
regimes: a stable one where events are isolated and non correlated which is 
characteristic of diffuse damage, and an unstable and dangerous one in which
events become synchronized and large icequakes are triggered.
\end{abstract}

\maketitle
\section{Introduction}

The fracturing of brittle heterogeneous material has often been studied at the
lab scale using acoustic emission measurements
\citep{Johansen&Sornette2000, Nechad&al2005}.
These studies reported an acceleration of brittle damage before failure.
Acoustic emission tools have already been used at meso-scale to find precursors to
natural gravity-driven instabilities such as cliff collapse
\citep{Amitrano&al2005} or slope instabilities
\citep{Dixon&Spriggs2007, Kolesnikov&al2003}. 
The present paper focuses on the acoustic emissions generated by an unstable glacier. 
To our knowledge, this is the first attempt to use these acoustic emissions
to predict the catastrophic collapse of a glacier.

Ice mass break-off is a natural gravity-driven instability as found in the
case of a landslide, rockfalls or
mountain collapse. Rupture takes place within ice and propagates until the
glacier collapses. This represents a considerable risk to mountain communities
situated below, especially in winter, as an ice avalanche may drag snow in
its train. 
An accurate prediction of this natural phenomenon must be made in order to
prevent such dangerous occurrences.
The first attempt to predict such icefalls was conducted in 1973 by
\citet{Flotron1977} and \citet{Roethlisberger1981a} on the Weisshorn hanging glacier. 
These authors measured the surface velocity of the unstable glacier and proposed an empirical function
to fit the increasing surface velocities before break-off. 
This function describes an acceleration of the surface displacement following
a power law up to infinity at a finite time $t_c$. 
Obviously, the real collapse will necessarily occur before $t_c$, but this
method gives a good description of the surface velocity evolution until rupture.
Recently, following \citet{Luethi2003} and \citet{Pralong&al2005}, \citet{Faillettaz&al2008} showed evidence of an oscillatory
behaviour superimposed on the general acceleration which enables a more
accurate determination of the time of rupture.
\citet{Faillettaz&al2008} showed also an increase in icequake activity before the
break-off. The aim of this paper is to present (i) a careful analysis of these seismic
measurements, (ii) our conclusions in terms of rupture processes, and (iii) perspectives
for forecasting.

Several studies have shown that glaciers can generate seismic signals called
icequakes.
Icequakes in Alpine glacier have been investigated in various
contexts. Several types of source mechanisms have been postulated or assumed
such as surface crevasse formation
\citep[e.g.][]{Neave&Savage1970,Deichmann&al2000}, stick-slip motion
\citep{Weaver&Malone1979,Roux&al2008}, bottom crevasse formation due to increased basal
drag during low subglacial water pressures \citep{Walter&al2008} or resonant water-filled cavities
\citep{Metaxian&al2003}.

In this study, the focus is on seismic activity generated by a cold hanging
glacier before its collapse. The crucial features of this type of glacier are:
(i) there is no sliding at the bedrock and (ii) there is no water within the ice. 
Precursory seismic signals were detected, and
a change in the behavior occurred two weeks before the global rupture.

\section{The Weisshorn glacier}

\begin{figure}
\noindent\includegraphics[width=20pc]{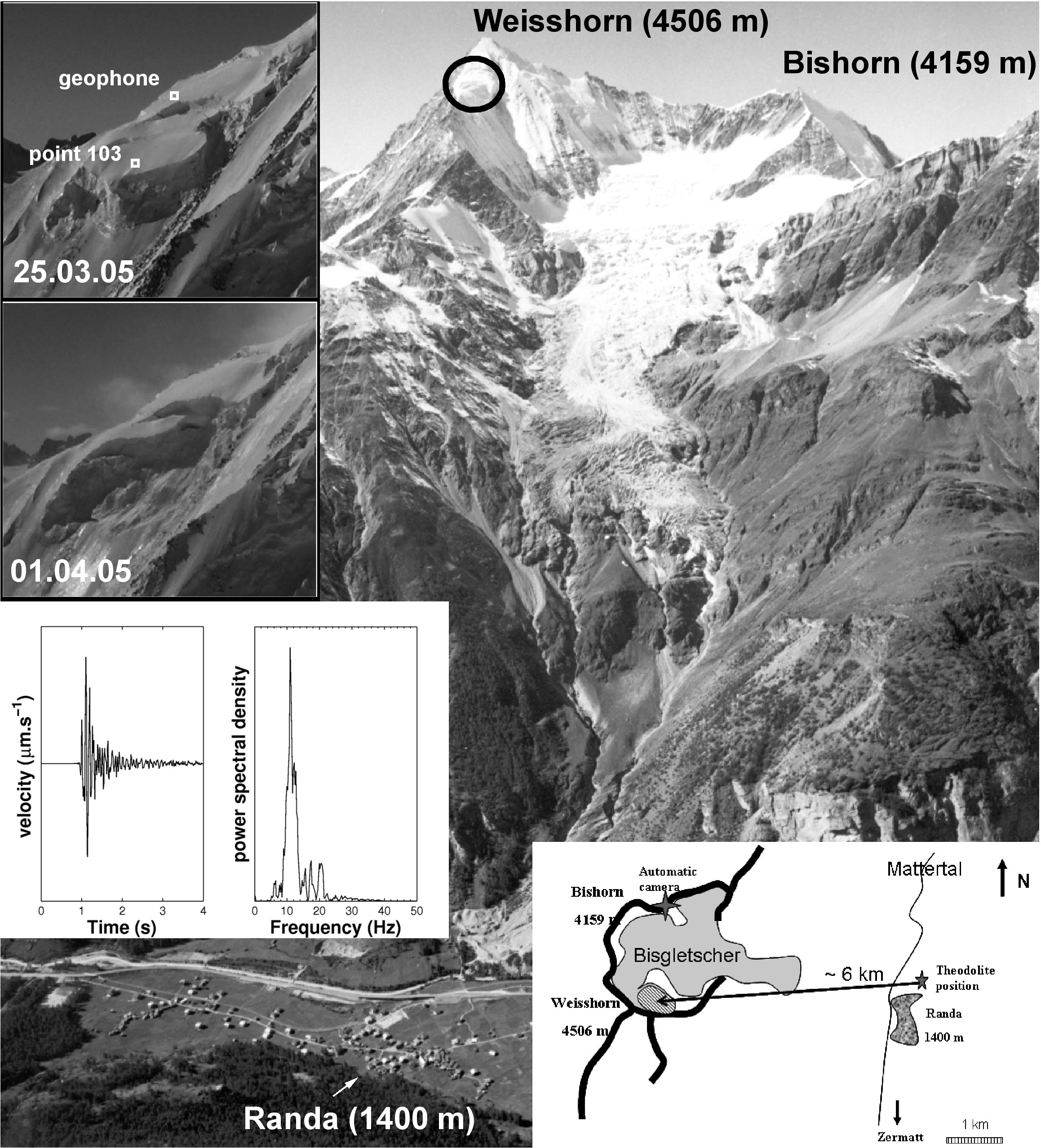}
\caption{\label{general}
The east face of Weisshorn with the hanging glacier. The
village of Randa and transit routes are visible in the valley. The ellipse
indicates the location of the hanging glacier. The left insets shows a closer
frontal view of the hanging glacier on March, 25th 2005 before the second break-off
(upper), and on April 1st, 2005 after the break-off (lower), including the positions
of the geophone and stake 103 used for
displacement measurements.
Below these photographs, unfiltered velocity seismogram of a typical
 event (maximum amplitude 2.5 $\rm \mu ms^{-1}$) and its corresponding
 normalized power spectrum density were drawn.
The bottom right inset shows the general schematic view of the Weisshorn hanging glacier (dashed zone), and
the monitoring setting (theodolite and automatic camera). Thick black lines
indicate the mountain ridges, and the thin line represents the bottom of the valley.}
\end{figure}

The northeast face of the Weisshorn (Valais, Switzerland) is covered with
unbalanced cold ramp glaciers \citep[i.e., the snow accumulation is, for the most part, compensated by
break-off,][]{Pralong&Funk2006} located between 4500 m and
3800 m a.s.l., on a steep slope of 45 to 50 degrees. In winter, snow
avalanches triggered by icefalls pose a recurrent threat to the village of
Randa located some 2500 m below the glacier, and to transit routes to
Zermatt (see Fig.~\ref{general}). 
The Weisshorn hanging glacier broke off five times in the last 35
years (1973, 1980, 1986, 1999 and 2005); two of these
events (in 1973 and 2005) were monitored in detail \citep{Flotron1977,Faillettaz&al2008}.

The total volume of the unstable ice mass was estimated at $0.5 \times 10^6
~\rm m^3$ by means of photogrammetry \citep{Faillettaz&al2008}.  
Because of the dangerous situation for the village of Randa, a
monitoring system was installed to alert the population of an impending
icefall.

An automatic camera (installed in  September 2003 on the Bishorn, see Fig.~\ref
{general}) provided a detailed movie of the destabilization of the
glacier. A first break-off occurred on March 24, 2005 (after
26.5 days of monitoring). Its estimated volume amounted to $100,000\;~\rm m^3$
(comparable to the 1973 break-off with $160,000\; ~\rm m^3$). On March 31,
2005 a second rupture occurred, during which the major part of the glacier
fell (after 33.5~days of monitoring). The volume of this second ice avalanche
was estimated at $400,000\; ~\rm m^3$. 

A geophone was also installed near the upper crevasse (see Fig.~\ref{general})
in order to record icequake activity before the final rupture. This signal is
assumed to describe the crack (or damage) evolution within the ice mass during the failure
process.

\section{Measurements and results}
A single geophone (Lennartz LE-3Dlite Mkll, 3 orthogonal sensors, with eigenfrequency of 1 Hz) was installed in firn
30~cm below the surface near the upper crevasse (Fig.~\ref{general}).
A Taurus portable seismograph (Nanometrics inc.) was used to record the seismic
activity of the glacier prior to its rupture with a sampling rate of 100 Hz. 
Unfortunately, the recorder failed on March 21, before the first break-off event, because of
battery problems.
A first seismic analysis of these measurements was presented in \citet{Faillettaz&al2008}.

\subsection{Signal characteristics}

The data show a high seismic emissivity from the hanging glacier during the time span of our observations.
\citet{Roux&al2008} found two types of signal characteristics for icequakes. The first type is
associated with small cracks and has a short and impulsive signal,
associated with crevasse opening. Such icequakes lie in the 10-40 Hz band and
therefore represent high-frequency events.
The second type of event is associated with long and complex signals with high
amplitude that can be linked with serac falls.

In the case of the Weisshorn hanging glacier, events with short and impulsive
signals with similar spectra are observed (see Fig. \ref{general}), with dominant
power contained in the 10-30 Hz frequency band. This observation is consistent with
previous results \citep{Roux&al2008,ONeel&al2007} 

This result is not surprising, as no serac falls could be observed during the time
span of our observations (based on daily photographs). 
Since the sensor is very close to the sources, attenuation is low.
The proximity of the source (less than 300 meters) gives rise to difficulties
in distinguishing P and S waves. 
As the geophone is situated above the upper crevasse separating the active and
the stable zone, compressive seismic waves
(primary wave) are perturbed by the discontinuity of the material and
therefore are less likely to be observed.

\subsection{Detection of events}

\begin{figure}
\noindent\includegraphics[width=20pc]{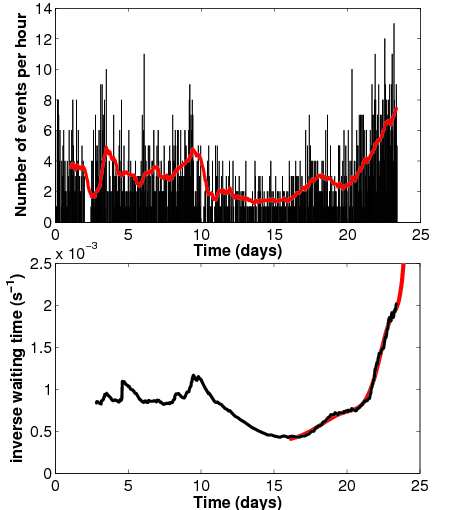}
\caption{\label{fig2} Top: Number of detected icequakes per hour (bars in black) as
 a function of time. In red is plotted the number of events per hour averaged
 on the last day of observation (sliding windows of 24 hours). Bottom:  Inverse of waiting time ($s^{-1}$) between two events as a function of time. Each point represents
the mean of the inverse of waiting time between 2 events in a moving
  window containing 200 events, plotted as a function of the time of the last point
of the window. In red is
 plotted the fit with the log-periodic law, indicating a critical time of 26.5
 days. The glacier broke-off after 26 days.
}
\end{figure}

The detection of events was performed in the following way. First, we evaluated
the root-mean-square (rms) of two concurrent time windows. The rms values over
the previous 800 ms long-term average (LTA) window and the previous 80 ms
short-term average (STA) windows were calculated and compared. When the ratio
$\gamma=\rm STA/LTA$ exceeded a given threshold (equal here to 3), an event was
detected and retained \citep{Allen1978,Walter&al2008}.
We found a total of 1731 such icequakes during the 21 days preceding the break-off.
Figure \ref{fig2} shows the number of detected events per hour during the 21-day period of the ice chunk destabilization.
An acceleration of the seismic activity was detected after 18 days of
monitoring (i.e., one week before the first and two weeks before the main
break-off). 

\subsection{Size-frequency distribution of icequake energies}

In order to compare all icequakes and perform a size-frequency analysis, we
first had to evaluate the size of each icequake.
Seismic event sizes were estimated based on their signal energies as defined for a
digitalized signal by (Evans [1979], in \citet{Amitrano&al2005})
\begin{equation}
E=\sum A^2\delta t 
\end{equation}
where $A$ is the signal amplitude and $\delta t$ the sampling period.
We made a manual selection of the beginning and the end of each signal and performed the
discrete summation for the evaluated duration.
This procedure allowed to determine the energy of every icequake.

\begin{figure}
\noindent\includegraphics[width=20pc]{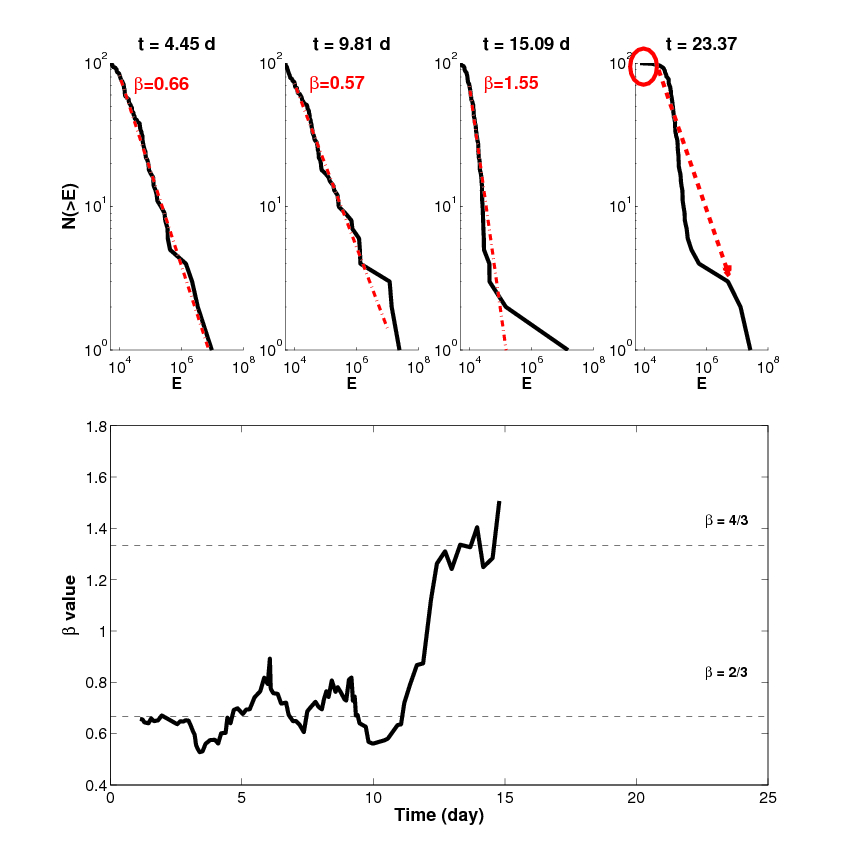}
\caption{\label{fig3}  
The four plots at the top show the complementary cumulative size-frequency distribution (CSFD) of
icequake energies obtained in four windows of 100 events each, ending at the
time indicated in the panels. The lower plot
shows the evolution of the exponent $\beta$ of the power law fitting the CSFD
obtained in running windows of 100 events. The exponent $\beta$
has been estimated with the Maximum
Likelihood method. After day 15, the CSFD cannot anymore be fitted
with a pure power law, because a significant ``shoulder'' appears in the tail.}
\end{figure}

The complementary cumulative size-frequency (also called ``survival'')  distribution (CSFD) of
the icequake energy preceding the break-off was then determined.
In order to study the temporal variation of the CSFD, we used a moving window
of 100 events with a 10-event shift
between successive windows. We analyzed the event size distribution
corresponding to each window (Fig. \ref{fig3} shows four typical windows).

Three different behaviors can be successively observed:

\noindent(i) For the windows located near the beginning of our
measurements (up to $t=10\;~\rm d$), the CSFD is well described by a power
law distribution over at least 2 orders of magnitude, indicating
a scale invariance of the acoustic emissions similar to that
characterizing earthquakes. For earthquakes, the corresponding
Gutenberg-Richter law is one of the most ubiquitous
statistical regularity observed  of earthquakes. It reads
\begin{equation}
\label{GR}
N(>E) \sim E^{-\beta}\;,
\end{equation}
where $N(>E)$ is the number of events of energy greater than $E$ and $\beta$ is 
the Gutenberg-Richter exponent found empirically close to $2/3$ for
shallow earthquakes (depths $< 70 km$) in subduction and transform fault
zones. Fig. \ref{fig3} shows the evolution of the exponent $\beta$ of the CSFD for
icequakes.
Up to $t=10$ d, the exponent $\beta$ is found compatible with the earthquake value $\beta =0.65 \pm 0.1$..

\noindent(ii) From $t=10\;~\rm d$ to $t=15 \;~\rm d$, the exponent
$\beta$ exhibits a rapid shift followed by a stabilization
to a value $\beta =1.4 \pm 0.1$, suggesting a change of
behavior of the damage process developing in the ice mass.
\citet{Pisarenko&Sornette2003} have associated a change of the exponent $\beta$
with a change in the rupture process: 
In their study, they confirmed previous observations that the exponent $\beta$ of the
Gutenberg-Richter law for shallow events is significantly smaller for subduction
zones ($\beta \simeq \frac{2}{3}$) compared to mid-ocean ridges ($\beta \simeq \frac{4}{3}$).
They proposed an explanation of these two exponent regimes based on two remarks.
First, large exponents $\beta$ are found in the distribution of acoustic emission energies
recorded for heterogeneous materials brought
to rupture, for which damage is mainly in the form of weak shear zones and open cracks.
In other words, large exponents $\beta$ characterize mainly an open crack mode of damage.
Second, when damage develops in the form of `` dislocations'' 
or mode II cracks, with slip mode of failure and with healing, 
the exponent $\beta$ is found smaller than $1$. This suggests that 
the low values of $\beta \simeq 2/3$ obtained for our data in the first $10$ days of recording
are associated with a
stable, slow and diffuse ``dislocation-like'' damage process. In the subsequent days, 
the increase of the exponent $\beta$ up to $1.4 \pm 0.1$ 
can be interpreted as revealing a transition to a mode of damage
controlled more and more by crack openings. 

\noindent(iii) For the time windows near the end of our observation period
(after $t=15 \;\rm d$), the CSFD of icequake energies
no longer conforms to a power law (upper
right panel of Fig. \ref{fig3}). 
The tail of the distribution develops a strong shoulder, indicating a change in the 
damage mechanism. The clear deficit of icequakes with small energies and the
excess of large ``characteristic'' events is fully compatible with the 
evolution of the second regime dominated by crack-like events which, by
their proliferation and fusion, progressively nucleate the run-away macro-crack
responsible for the final avalanche associated with a rather clean
crack-like rupture, as shown in Fig. 1.

\subsection{Accelerating rate of icequakes}
The time evolution of the rate of icequakes is well-captured by that of
the inverse of the mean time lag between two consecutive
icequakes. Fig. \ref{fig2} shows this inverse mean time lag
in a moving window containing 100 events as a function of the time of the last point of
the window. One can clearly observe a general acceleration of the  icequake
activity a few days before the collapse of the glacier.
The beginning of this acceleration observed for $t>15\;~\rm d$ 
coincides approximately with the transition to the crack-dominated damage
regime, that we have identified from 
the increase of the exponent $\beta$ shown in Fig. 3.

In addition to its acceleration developing in the last few days preceding the collapse,
the rate $s(t)$ of icequakes is also characterized by accelerating oscillations,
which are well-fitted by the log-periodic power law
defined by the equation:
\begin{equation}
\label{logper}
s(t)=s_0+a(t_c-t)^{m}\Big[1+C\sin(2\pi\frac{\ln(t_c-t)}{\ln(\lambda)}+D)\Big]~.
\end{equation}
Here, $s_0$ is a constant, $t_c$ is the critical time at which the 
global collapse is expected, $m<1$ is the power
law exponent quantifying the acceleration, $a$ is a constant, $C$ is the relative amplitude
of the oscillations with respect to the overall power law acceleration, $\lambda$ is the
so-called ``scaling ratio'' associated with the log-periodicity of expression (\ref{logper})
and $D$ is the phase of the log-periodic oscillation. The best fit of the rate of icequakes
to expression (\ref{logper}) is shown in red in Fig. \ref{fig2}. It predicts a critical time $t_c$
very close (half-a-day) to the time when the glacier broke-off, suggesting a novel
method to predict future glacier failures by monitoring icequake rates in real time.

Such log-periodic power law has been found to be an efficient practical 
tool to model acoustic emission data (see \cite{Johansen&Sornette2000} and references
therein). 
The log-periodic oscillations in particular
provide an important additional information to stabilize the fits
and get better estimations of the critical time $t_c$. 
Log-periodic oscillations are the signature of a weaker kind of scale invariance
called discrete scale invariance, appearing after the partial breaking of a
continuous scale symmetry. In quasi-static growth models of ensembles
of interacting cracks, discrete scale invariance has been shown to originate
from the existence of a cascade of Mullins-Sekerka instabilities
\citep{Huang&al1997}. This view is
coherent with the results obtained in the previous
section in which we have argued that, for $t>15\;~\rm d$,
crack growth and coalescence dominate, developing into 
a collective crack damage mode leading
eventually to the global collapse. Furthermore, the log-periodic power law
model was previously found to provide a good description of the surface motion of
different points on this glacier \citep{Faillettaz&al2008}, showing the 
convergent evidence offered by these two metrics (icequake rate and surface deformation).

\section{Summary and prospective}

We have presented and studied a unique dataset of icequakes recorded
in the immediate vicinity of the hanging Weisshorn glacier over 21 days prior to its rupture.
While measurement records ceased unfortunately 3 days before the first
break-off and 10 days before the larger subsequent one, we
have nevertheless been able to obtain a coherent quantitative picture of the 
damage process developing before the impending glacier collapse.
Our main results include the demonstration of
(i) a clear increase of the icequake activity within the glacier (measured as the
inverse of the waiting time between successive icequakes) 
starting approximately $7$ days before the first avalanche, (ii) a two-step evolution of the
size-frequency distribution of icequake energies, characterizing
a first transition to a crack-like dominated damage followed by a second
transition in which large characteristic cracks are thought to prepare the nucleation
of the run-away rupture, and
(iii) a log-periodic oscillatory behavior superimposed on a power law
acceleration of the rate of icequakes, which is a typical of the hierarchical
cascade of rupture instabilities found in earlier reports on the acoustic
emissions associated with the failure of heterogeneous materials.

Provided that technical solutions are found to ensure continuous
icequake recordings in the difficult high-altitude mountain 
conditions, our results open the road for real-time diagnostics
of impending glacier failure. The next steps towards this goal include
(a) developing an automatic seismic data processing in real time
(which include the automatic detection of icequakes and the determination
of their energy), (b) processing these data with the statistical tools
developed here and (c) performing systematic reliability tests to access
the rate of false alarms (false positives or errors of type I) versus missed 
events (false negatives or errors of type II). Step (c) is necessary
for an informed cost-benefit analysis of the societal and economic
impacts of the proposed real-time forecast methodology.


{\bf Acknowledgements}:
The Institute of Geophysics at the ETHZ is gratefully acknowledged for
allowing us the use of their instruments.
Thanks are extended to Fabian Walter for fruitful
discussions. 


\vskip -0.7cm
\begin{thebibliography}{}
\bibliographystyle{igs}
\expandafter\ifx\csname natexlab\endcsname\relax
 \def\natexlab#1{#1}\fi
\expandafter\ifx\csname selectlanguage\endcsname\relax
 \def\selectlanguage#1{\relax}\fi


\bibitem[Allen(1978)]{Allen1978}
Allen, R. V., 1978.
Automatic earthquake recognition and timing from single traces.
{\em Bull. Seismol. Soc. Am.}, {\bf 68}, (5), 1521-1532.

\bibitem[Amitrano et al.(2005)]{Amitrano&al2005}
Amitrano, D., Grasso J.-R. and Senfaute, G., 2005.
Seismic precursory patterns before a cliff collapse and critical point phenomena,
{\em Geophys. Res. Lett}, {\bf 32}:L08314.









\bibitem[Deichmann et al.(2000)]{Deichmann&al2000}
Deichmann, N., Ansorge, J., Scherbaum, F., Aschwanden, A.,
Bernhardi, F. and Gudmundsson, G. H., 2000.
Evidence for deep icequakes in an Alpine glacier,
{\em Ann. Glaciol.}, {\bf 31}, 85-90.


\bibitem[Dixon and Spriggs(2007)]{Dixon&Spriggs2007}
Dixon, N. and Spriggs, M., 2007.
Quantification of slope displacement rates using acoustic emission monitoring,
{\em Can. Geotech. J.}, {\bf 44}, 966-976.



\bibitem[Faillettaz et al.(2008)]{Faillettaz&al2008}
Faillettaz, J., Pralong, A., Funk, M. and Deichmann, N., 2008.
Evidence of log-periodic oscillations and increasing icequake activity during
the breaking-off of large ice masses.
{\em J. Glaciol.}, {\bf 57}, (187), 725. 

\bibitem[Flotron(1977)]{Flotron1977}
Flotron, A., 1977.
Movement studies on hanging glaciers in relation with an ice avalanche,
{\em J. Glaciol.}, {\bf 19} (81), 671--672.

\bibitem[Huang et al.(1997)]{Huang&al1997}
Huang, Y., Ouillon, G., Saleur H. and Sornette, D., 1997.
Spontaneous generation of discrete scale invariance in growth models,
{\em Phys. Rev. E}, {\bf 55} (6), 6433-6447.




\bibitem[Johansen and Sornette(2000)]{Johansen&Sornette2000}
Johansen, A. and Sornette, D., 2000.
Critical ruptures.
{\em Eur. Phys. J. B}, {\bf 18}, 163.

\bibitem[Kolesnikov et al.(2003)]{Kolesnikov&al2003}
Kolesnikov, Yu. I., Nemirovich-Danchenko, M. M., Goldin, S. V. and Seleznev
V. S., 2003.
Slope stability monitoring from microseismic field using polarization
methodology.
{\em Nat. Haz. Earth Sys. Sc.}, {\bf 3}, 515-521.

\bibitem[L\"uthi(2003)]{Luethi2003}
L{\"u}thi, M., 2003.
Instability in glacial systems,
Milestones in Physical Glaciology: From the Pioneers to a Modern Science,
180, 63--70, VAW, ETH-Z\"urich.

\bibitem[Metaxian et al.(2003)]{Metaxian&al2003}
M\'etaxian, J.P., Araujo, S., Mora, M. and Lesage, P., 2003.
Seismicity related to the glacier of Cotopaxi Volcano, Ecuador,
{\em Geophys. Res. Lett.}, {\bf 30} (9), 1483.

\bibitem[Neave and Savage(1970)]{Neave&Savage1970}
Neave, K. G. and Savage, J. C., 1970.
Icequakes at Athabasca Glacier.
{\em J. Glaciol.}, {\bf 49}, 587-598.

\bibitem[Nechad et al.(2005)]{Nechad&al2005}
Nechad, H.,  Helmstetter, A., El Guerjouma, R. and Sornette, D., 2005.
Andrade and Critical Time-to-Failure Laws in Fiber-Matrix Composites:
Experiments and Model,  
{\em J. Mech. Phys. Solids}, {\bf 53}, 1099-1127.

\bibitem[O'Neel et al.(2007)]{ONeel&al2007}
O'Neel, S., Marshall, H. P., McNamara, D.E. and Pfeffer, W. T., 2007.
Seismic detection and analysis of icequakes at Columbia Glacier, Alaska.
{\em J. Geophys. Res.}, {\bf 112}, F03S23.

\bibitem[Pisarenko and Sornette(2003)]{Pisarenko&Sornette2003}
Pisarenko, V.F. and Sornette, D., 2003.
Characterization of the frequency of extreme earthquake events by generalized
Pareto distribution,
{\em Pure Appl. Geophys.}, {\bf 160}, 2343-2364.



\bibitem[Pralong and Funk(2006)]{Pralong&Funk2006}
Pralong, A. and M. Funk, M., 2006.
On the instability of avalanching glaciers,
{\em J. Glaciol.}, {\bf 52} (176), 31--48.


\bibitem[Pralong et al.(2005)]{Pralong&al2005}
Pralong, A., Birrer, C., Stahel, W. and Funk, M., 2005.
On the Predictability of Ice Avalanches,
{\em Nonlin. Processes Geophys.}, {\bf 12}, 849--861.



\bibitem[R\"othlisberger(1981)]{Roethlisberger1981a}
R\"othlisberger, H., 1981.
Eislawinen und Ausbr{\"u}che von Gletscherseen,
in P. Kasser ({E}d.), Gletscher und Klima - glaciers et climat, 
Jahrbuch der Schweizerischen Naturforschenden Gesellschaft, wissenschaftlicher
Teil 1978,pp 170--212,
Birkh\"auser Verlag Basel, Boston, Stuttgart.



\bibitem[Roux et al.(2008)]{Roux&al2008}
Roux, P.-F., Marsan, D., Metaxian, J-P., O'Brien, G. and Moreau L., 2008.
Microseismic activity within a serac zone in an alpine glacier (Glacier
d'Argenti\`ere, Mont-Blanc, France).
{\em J. Glaciol.},{\bf 54}, (184), 157.














\bibitem[Walter et al.(2008)]{Walter&al2008}
Walter, F., Deichmann, N. and Funk, M., 2008.
Basal icequakes during changing subglacial water pressures beneath
Gornergletscher, Switzerland.
{\em J. Glaciol.}, {\bf 54}, (186), 511.

\bibitem[Weaver and Malone(1979)]{Weaver&Malone1979}
Weaver, C. and Malone, S., 1979.
Seismic evidence for discrete glacier motion at the rock-ice interface.
{\em J. Glaciol.}, {\bf 23}, (89), 171.



\bibliography{/home/jeromef/glacier.bib}
\end{thebibliography}
\end{document}